\documentclass[reprint, longbibliography]{iopjournal}
\usepackage{amsmath}
\usepackage{amssymb}
\usepackage{comment}
\usepackage[caption=false]{subfig} 
\usepackage{pifont}
\usepackage{accents} 

\usepackage{dcolumn}
\usepackage{bm}
\usepackage{cite}

\begin{document}

\articletype{paper} 

\title{Odd transport in a two-temperature Brownian dimer}

\author{Iman Abdoli$^{1, *}$\orcid{0000-0002-2309-8560} and Hartmut L\"{o}wen$^{1}$\orcid{0000-0001-5376-8062}}

\affil{$^1$Institut für Theoretische Physik II - Weiche Materie, Heinrich-Heine-Universität Düsseldorf, Universitätsstraße 1, D-40225 Düsseldorf, Germany}


\affil{$^*$Author to whom any correspondence should be addressed.}

\email{Iman.abdoli@hhu.de}


\begin{abstract}
	
We investigate a two-temperature Brownian dimer with odd mobility, characterized by antisymmetric transport coefficients, as a controlled paradigm for odd nonequilibrium dynamics. The system is made of two harmonically confined particles coupled by an elastic spring and connected to reservoirs at different temperatures. Odd mobility converts conservative forces into transverse motion, linking heat exchange to circulating probability currents without requiring external torques, spatial anisotropy, or nonconservative driving. Our exact solution shows that odd mobility creates handed correlations between the two particles while leaving the individual particle distributions isotropic. These correlations arise only when temperature imbalance, elastic coupling, and odd mobility act together, and their handedness reverses when the odd response is reversed. The steady probability current contains two distinct parts: the ordinary irreversible current of a two-temperature dimer and an additional handed contribution generated by odd mobility. When projected onto the motion of each particle, this handed contribution becomes a pair of counter-rotating circulating currents inside the traps. Based on the currents we compute the heat transfer and entropy production analytically. We show that odd mobility enhances thermal conductance between the reservoirs, while the net heat current and total dissipation remain unchanged under reversal of the odd handedness.


\end{abstract}

\section{Introduction}
\vspace{5mm}

Odd transport has emerged as a symmetry-based framework for systems in which the usual reciprocal form of linear response is relaxed, allowing transport tensors to acquire antisymmetric components~\cite{fruchart2023odd}. For an overdamped Brownian particle in equilibrium, isotropy and detailed balance reduce the mobility tensor to a scalar coefficient, so that the mean velocity is parallel to the applied force. When reciprocity is broken, for example by microscopic chirality, gyrotropic dynamics, or time-reversal-symmetry breaking, a transverse force response is also permitted. In two spatial dimensions, the most general rotationally invariant mobility tensor takes the form
\begin{equation}
	\boldsymbol{\mu} = \mu_0(\mathbf I + \kappa \boldsymbol{\varepsilon}),
	\label{eq:intro_mobility}
\end{equation}
where \(\mu_0\) is the bare mobility, \(\mathbf I\) is the identity tensor, \(\boldsymbol{\varepsilon}\) is the two-dimensional Levi--Civita tensor, and \(\kappa\) controls the strength and handedness of the transverse response. Such a transport law generates Hall-like fluxes in addition to ordinary force-parallel or down-gradient motion, while remaining compatible with rotational isotropy~\cite{hargus2021odd,poggioli2023odd,faedi2026mobility,wojcik2026chiral}. In homogeneous odd diffusion, the antisymmetric diffusive flux is divergence-free in the bulk; consequently, odd transport can leave steady density profiles unchanged while strongly modifying probability currents, relaxation pathways, and boundary-induced transport~\cite{vuijk2019anomalous,abdoli2020nondiffusive,vega2022diffusive,abdoli2026dynamical, caprini2025active}.

The same symmetry principle appears across a wide range of nonequilibrium systems. Passive realizations include charged Brownian particles in external magnetic fields, where Lorentz forces generate effective transverse fluxes and odd-diffusive dynamics~\cite{czopnik2001brownian,chun2018emergence,abdoli2020stationary,abdoli2023odd,wittmann2025confined}, and magnetic skyrmions, whose gyrotropic motion produces Hall-like drift and diffusion~\cite{schutte2014inertia,troncoso2014brownian,reichhardt2015collective,litzius2017skyrmion,fert2017magnetic,buttner2018theory,weissenhofer2021skyrmion}. Active and biological systems provide further examples through intrinsic rotation, circular swimming, internal spin, and vortical collective motion~\cite{van2008dynamics,kummel2013circular,diluzio2005escherichia,marcos2012bacterial,petroff2015fast,drescher2009dancing,muzzeddu2022active,caprini2025selfwrapping,marini2026emergent}. At continuum and solid-like scales, antisymmetric response appears as odd viscosity, odd elasticity, and odd viscoelasticity~\cite{banerjee2017odd,han2021fluctuating,markovich2021odd,scheibner2020odd,braverman2021topological,banerjee2021active,lier2022passive,huang2025oddcrystals}. In interacting or confined systems, these transverse responses can produce interaction-enhanced diffusion, oscillatory force correlations, reversed density wakes, negative mobility, edge and interface currents, and modified collective relaxation~\cite{kalz2022collisions,kalz2024oscillatory,luigi2025self,faedi2026mobility,soni2019odd,massana2021arrested,bililign2022motile,tan2022odd,mecke2024emergent,goerlich2026particle,abdoli2026dynamical}. Thus odd transport provides a route to nonequilibrium currents and correlations without changing the underlying conservative potential landscape.

A complementary route to nonequilibrium steady states is obtained when different components of a Brownian system are coupled to different thermal reservoirs. In this case, the drive is assigned by species: each component experiences its own effective temperature or diffusivity, and irreversibility arises through their coupling~\cite{grosberg2015nonequilibrium}. This setting has been realized experimentally with two optically trapped Brownian beads, where random forcing of one trap creates an effective temperature difference and produces measurable energy transfer, modified variances, and interparticle cross-correlations~\cite{berut2014energy,berut2016stationary}. Planar scalar-mobility dimers further show that coupling two particles at different temperatures can generate circulating probability currents and synchronized current rotations in configuration space~\cite{dotsenko2023out}. At the many-body level, differential-diffusivity mixtures provide a minimal Brownian analogue of active--passive mixtures~\cite{stenhammar2015activity}: they can demix into cold dense clusters and hot dilute regions~\cite{weber2016binary,ilker2020phase}, although this tendency becomes re-entrant at high density when phase separation no longer lowers entropy production relative to the mixed state~\cite{mccarthy2024demixing}.

These two lines of work raise a natural question: how is two-temperature Brownian transport modified when the microscopic response itself is odd? Existing dimers and hot--cold mixtures mostly assume scalar mobility, so that circulation arises from coupling, temperature imbalance, and geometry. Here we remove this restriction in the simplest analytically solvable setting: two overdamped particles in two dimensions, confined by isotropic harmonic traps, elastically coupled, and connected to reservoirs at temperatures \(T_1\) and \(T_2\). The particles have the rotationally invariant odd mobility in Eq.~\eqref{eq:intro_mobility}; hence each conservative force produces both longitudinal and transverse drift. Because the traps and the coupling are isotropic, the model contains no imposed torque or spatial anisotropy, isolating the effect of odd mobility on heat exchange and irreversible currents. The exact solution shows that the steady state remains Gaussian but acquires a transverse interparticle correlation, \(\langle x_1 y_2\rangle=-\langle y_1 x_2\rangle\), which exists only when temperature imbalance, elastic coupling, and odd mobility are all present. We show that the same odd response generates handed current components and counter-rotating marginal circulations. From these currents we derive the heat exchanged with the two reservoirs and find an exact expression for the entropy production rate. In this effective odd-transport model, the odd response enhances thermal conductance between the baths, while the net heat flow remains unchanged when the sign of the odd parameter is flipped. The model therefore distinguishes odd observables, such as transverse correlations and handed current components, from scalar thermodynamic observables, which are insensitive to the sign of the odd response.

\section{Model}
\label{sec:model}
\vspace{5mm}
\begin{figure}[tbp]
	\centering
	\includegraphics[width=0.8\textwidth]{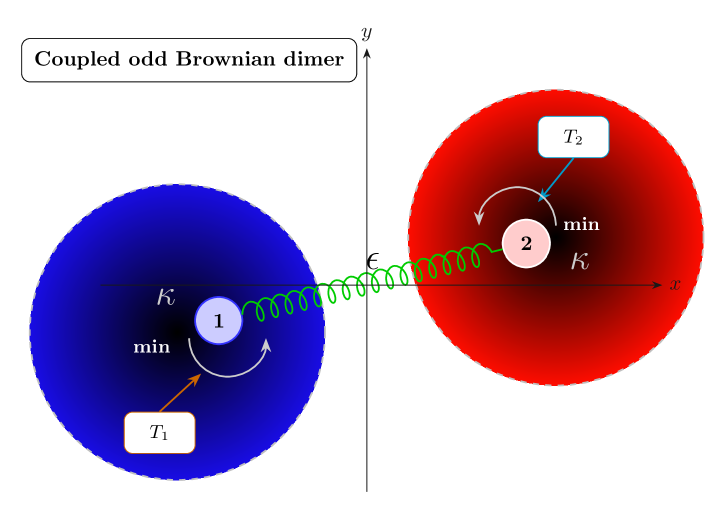}
	\caption{
		Coupled odd Brownian dimer. 
		Two overdamped Brownian particles are confined by harmonic traps of stiffness \(k\) and coupled by a harmonic spring of stiffness \(\epsilon\). 
		The colored disks schematically represent the harmonic confining potentials, with the darker central regions marking the trap minima. The blue disk denotes particle \(1\), coupled to the colder reservoir \(T_1\), while the red disk denotes particle \(2\), coupled to the hotter reservoir \(T_2>T_1\).
		Odd transport is encoded in the mobility tensor \(\boldsymbol{\mu}=\mu_0(\mathbf I+\kappa\boldsymbol{\varepsilon})\), so that a force generates both a longitudinal response and a transverse response whose handedness is set by the sign of \(\kappa\). 
		The combination of temperature imbalance, interparticle coupling, and odd mobility produces chiral correlations, circulating probability currents, and heat transfer between the reservoirs.
	}
	\label{fig:odd_dimer}
\end{figure}

We consider two overdamped Brownian particles in two dimensions, with positions \(\mathbf r_i=(x_i,y_i)^{\mathrm T}\) for \(i=1,2\), as illustrated in Fig.~\ref{fig:odd_dimer}. Each particle is confined by an isotropic harmonic trap of stiffness \(k\), and the two are coupled by a spring of stiffness \(\epsilon\). The total potential energy is
\begin{equation}
	U(\mathbf r_1,\mathbf r_2)
	=
	\frac{k}{2}\left(|\mathbf r_1|^2+|\mathbf r_2|^2\right)
	+
	\frac{\epsilon}{2}|\mathbf r_1-\mathbf r_2|^2 .
	\label{eq:model_potential}
\end{equation}
The corresponding forces are \(\mathbf F_1=-(k+\epsilon)\mathbf r_1+\epsilon\mathbf r_2\) and \(\mathbf F_2=-(k+\epsilon)\mathbf r_2+\epsilon\mathbf r_1\). The potential is invariant under simultaneous rotations of \(\mathbf r_1\) and \(\mathbf r_2\), so it contains no spatial anisotropy, external torque, or chiral bias.
Odd transport enters through the mobility tensor \(\boldsymbol{\mu}=\mu_0(\mathbf I+\kappa\boldsymbol{\varepsilon})\), so that the deterministic drift of particle \(i\) is \(\boldsymbol{\mu}\mathbf F_i\). The same force thus generates both a longitudinal response and a transverse response controlled by the oddness parameter \(\kappa\); the sign of \(\kappa\) sets the handedness, and \(\kappa=0\) recovers the scalar-mobility dimer. The particles are connected to reservoirs at temperatures \(T_1\) and \(T_2\), with \(T_2>T_1\) and the Boltzmann constant  \(k_{\mathrm B}=1\). Apart from their temperatures, the two particles are mechanically identical: they share the same \(k\), \(\epsilon\), and \(\boldsymbol{\mu}\). The temperature difference enters through the reservoir-dependent transport tensors \(T_i\boldsymbol{\mu}\), while the spring provides the channel for heat, correlations, and probability currents between the particles.

We describe the dynamics at the level of the joint probability density of finding particle 1 at \((x_1,y_1)\) and particle 2 at \((x_2,y_2)\) at time \(t\), \(P(\mathbf R,t)\), where \(\mathbf R=(x_1,y_1,x_2,y_2)^{\mathrm T}\). Its evolution can be described by the following continuity equation
\begin{equation}
	\frac{\partial P(\mathbf R,t)}{\partial t}  = -\nabla_{\mathbf R}\cdot \mathbf J(\mathbf R,t),
	\label{eq:model_continuity}
\end{equation}
where \(\mathbf J(\mathbf R,t)\) is the probability current in configuration space. The configuration-space current is written as
\(\mathbf J=(\mathbf J_1,\mathbf J_2)^{\mathrm T},\)
where \(\mathbf J_i\) is the two-dimensional probability current associated with particle \(i\). Explicitly
\begin{equation}
	\mathbf J_i
	=
	-\boldsymbol{\mu}(\nabla_i U)P
	-
	T_i\boldsymbol{\mu}\nabla_i P,
	\label{eq:fluxes}
\end{equation}
where \(\nabla_i=(\partial_{x_i},\partial_{y_i})^{\mathrm T}\). The first term is the drift current generated by the conservative force, while the second term is the reservoir-dependent diffusive current.


Since the potential in Eq.~\eqref{eq:model_potential} is quadratic, it can be written as \(U=\frac12\mathbf R^{\mathrm T}\mathbf K\mathbf R\), with the block matrix
\begin{equation}
	\mathbf K
	=
	\begin{pmatrix}
		(k+\epsilon)\mathbf I & -\epsilon\mathbf I \\
		-\epsilon\mathbf I & (k+\epsilon)\mathbf I
	\end{pmatrix}.
	\label{eq:model_K}
\end{equation}
The corresponding current can be written in the compact matrix form as
\begin{equation}
	\mathbf J
	=
	-\mathbf A\mathbf R\,P
	-
	\mathbf D\nabla_{\mathbf R}P,
	\qquad \text{with} \qquad 
	\mathbf A=\boldsymbol{\mathcal M}\mathbf K,
	\qquad
	\mathbf D=
	\begin{pmatrix}
		T_1\boldsymbol{\mu} & \mathbf 0 \\
		\mathbf 0 & T_2\boldsymbol{\mu}
	\end{pmatrix},
	\label{eq:model_current}
\end{equation}
where \(\mathbf 0\) is the two-dimensional zero matrix  and \(\boldsymbol{\mathcal M}=\mathrm{diag}(\boldsymbol{\mu},\boldsymbol{\mu})\) is the four-dimensional mobility matrix. The drift matrix \(\mathbf A\) describes relaxation in the harmonic potential modified by the odd mobility. The matrix \(\mathbf D\) is the temperature-weighted transport matrix: its symmetric part sets the noise covariance, while its antisymmetric part, proportional to \(\kappa\), contributes to the current without affecting the second-order diffusion operator. Oddness thus enters both the deterministic drift and the diffusive part of the current, yet the potential itself remains conservative.

Equations~\eqref{eq:model_continuity} and \eqref{eq:model_current} define a linear Fokker--Planck problem with constant coefficients. The system is controlled by three dimensionless parameters: the temperature ratio \(\Theta=T_2/T_1\), the relative coupling \(\tilde\epsilon=\epsilon/k\), and the oddness parameter \(\kappa\). These govern the exact steady state, its odd correlations, and the irreversible currents analyzed below.

\section{Exact nonequilibrium steady state and odd correlations}
\label{sec:steady_state_odd_correlations}
\vspace{5mm}
\begin{figure}[tbp]
	\centering
	\includegraphics[width=1\textwidth]{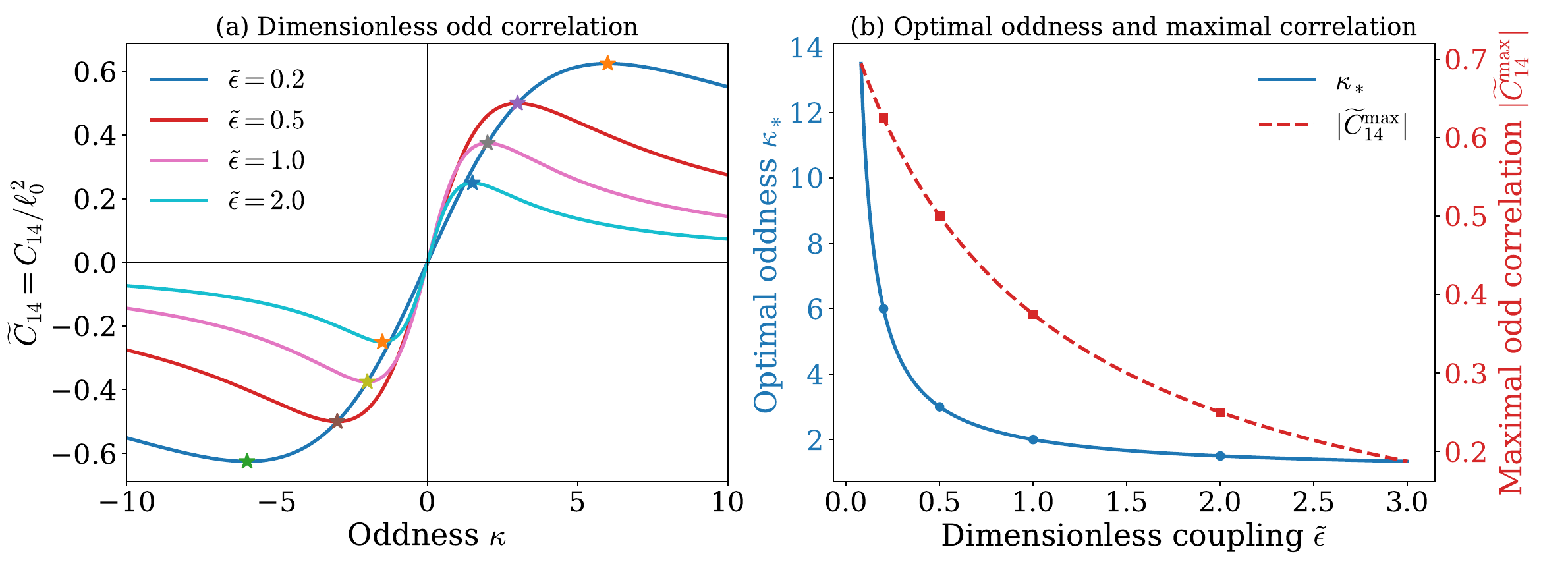}
	\caption{
		Dimensionless odd interparticle correlation and its optimal oddness for the temperature ratio \(\Theta=T_2/T_1=4.0\). 
		\textbf{(a)} The cross-correlation \(\widetilde C_{14}=C_{14}/\ell_0^2=\langle x_1y_2\rangle/\ell_0^2\) from Eq.\eqref{eq:ss_C14_dimensionless} as a function of the oddness parameter \(\kappa\), for different dimensionless coupling strengths \(\tilde\epsilon=\epsilon/k\). Here \(\ell_0=\sqrt{T_1/k}\) is the characteristic length scale.  
		The correlation is odd in \(\kappa\), vanishes at \(\kappa=0\), and changes sign when the handedness of the odd mobility is reversed. 
		Stars mark the extrema at \(\kappa=\pm\kappa_\ast\), showing that the odd correlation is nonmonotonic and is maximized at a finite value of the oddness. 
		\textbf{(b)} Optimal oddness \(\kappa_\ast=(1+\tilde\epsilon)/\tilde\epsilon\) and maximal correlation magnitude \(|\widetilde C_{14}^{\max}|=(\Theta-1)/[4(1+\tilde\epsilon)]\). 
		Weak coupling shifts the optimum to large \(|\kappa|\), whereas stronger coupling moves the optimum toward \(|\kappa|\simeq1\) but reduces the maximal dimensionless odd correlation. 
		The odd cross-correlation therefore requires the simultaneous presence of thermal nonequilibrium, interparticle coupling, and odd mobility.
	}
	\label{fig:odd_correlation}
\end{figure}

Because the drift is linear in the configuration vector \(\mathbf R\) and the transport matrices in Eqs.~\eqref{eq:model_continuity} and \eqref{eq:model_current} are constant, the steady state can be obtained exactly. In the steady state, the probability current is divergence-free, \(\nabla_{\mathbf R}\cdot\mathbf J=0\), but it need not vanish. The stationary probability density is therefore Gaussian and can be written as
\begin{equation}
	P_{\mathrm{ss}}(\mathbf R)
	=
	\frac{1}{(2\pi)^2\sqrt{\det\mathbf C}}
	\exp\left[
	-\frac{1}{2}\mathbf R^{\mathrm T}\mathbf C^{-1}\mathbf R
	\right],
	\label{eq:ss_gaussian}
\end{equation}
where \(\mathbf C=\langle \mathbf R\mathbf R^{\mathrm T}\rangle_{\mathrm{ss}}\) is the steady-state covariance matrix. The covariance is determined by the Lyapunov equation~\cite{abdoli2020correlations, abdoli2022tunable}
\begin{equation}
	\mathbf A\mathbf C+\mathbf C\mathbf A^{\mathrm T}
	=
	2\mathbf D_s ,
	\label{eq:ss_lyapunov}
\end{equation}
where \(\mathbf A\) is the linear drift matrix and \(\mathbf D_s=(\mathbf D+\mathbf D^{\mathrm T})/2\) is the symmetric part of the transport matrix, which are defined in Eq.~\eqref{eq:model_current}. The explicit matrices and the solution of Eq.~\eqref{eq:ss_lyapunov} are given in ~\ref{app:steady_state_derivation}.

The covariance matrix has the symmetry-constrained form
\begin{equation}
	\mathbf C
	=
	\begin{pmatrix}
		C_{11} & 0 & C_{13} & C_{14} \\
		0 & C_{11} & C_{23} & C_{13} \\
		C_{13} & C_{23} & C_{33} & 0 \\
		C_{14} & C_{13} & 0 & C_{33}
	\end{pmatrix}.
	\label{eq:ss_covariance_structure}
\end{equation}

The diagonal entries \(C_{11}=\langle x_1^2\rangle=\langle y_1^2\rangle\) and \(C_{33}=\langle x_2^2\rangle=\langle y_2^2\rangle\) are the one-particle variances of particles \(1\) and \(2\), respectively. The ordinary interparticle correlation is \(C_{13}=\langle x_1x_2\rangle=\langle y_1y_2\rangle\), whereas the transverse correlations \(C_{14}=\langle x_1y_2\rangle=-C_{23}\) encode the odd structural part of the steady state. 

We introduce the length scale \(\ell_0=\sqrt{T_1/k}\). The marginal widths can then be written as \(C_{11}=\ell_0^2\sigma_1^2\) and \(C_{33}=\ell_0^2\sigma_2^2\) with


\begin{equation}
	\sigma_1^2
	=
	\frac{
		(1+\tilde\epsilon)
		\left[
		2+4\tilde\epsilon
		+
		(1+\kappa^2)\tilde\epsilon^2(1+\Theta)
		\right]
	}
	{
		2(1+2\tilde\epsilon)Q_\kappa
	},
	\label{eq:ss_sigma1}
\end{equation}
and
\begin{equation}
	\sigma_2^2
	=
	\frac{
		(1+\tilde\epsilon)
		\left[
		2\Theta+4\tilde\epsilon\Theta
		+
		(1+\kappa^2)\tilde\epsilon^2(1+\Theta)
		\right]
	}
	{
		2(1+2\tilde\epsilon)Q_\kappa
	}.
	\label{eq:ss_sigma2}
\end{equation}
where \(Q_\kappa=(1+\tilde\epsilon)^2+\kappa^2\tilde\epsilon^2\) and \(\Theta=T_2/T_1\).
Their difference takes the simple form
\begin{equation}
	C_{33}-C_{11}
	=
	\ell_0^2
	\frac{(1+\tilde\epsilon)(\Theta-1)}{Q_\kappa}.
	\label{eq:ss_width_difference}
\end{equation}
This difference will be relevant for our subsequent study of the marginal probability densities. 

The genuinely odd structural signature is contained in the transverse interparticle correlation. This quantity measures how a displacement of particle \(1\) along the \(x\)-direction is correlated with a displacement of particle \(2\) along the perpendicular \(y\)-direction. It is therefore different from the ordinary positional correlation \(C_{13}\), which correlates parallel displacements of the two particles. Together with \(C_{23}=-C_{14}\), it forms an antisymmetric cross-covariance between the two particles: a positive fluctuation of one particle in one direction is statistically accompanied by a transverse fluctuation of the other particle. Thus \(C_{14}\) quantifies the handed tilt of the steady-state Gaussian in the full configuration space, while leaving the one-particle marginals isotropic. In dimensionless form
\begin{equation}
	\widetilde C_{14}
	=
	\frac{C_{14}}{\ell_0^2}
	=
	\frac{
		\tilde\epsilon\kappa(\Theta-1)
	}
	{
		2Q_\kappa
	}.
	\label{eq:ss_C14_dimensionless}
\end{equation}
This expression shows that the transverse covariance is not produced by temperature imbalance or coupling alone. It appears only when the thermal drive is transmitted through the elastic coupling in the presence of odd mobility. Its sign is set by the handedness of the odd response, whereas its magnitude is controlled by the competition between coupling strength and odd transverse mobility.


The dimensionless odd correlation \(\widetilde C_{14}\) is antisymmetric under \(\kappa\to-\kappa\), reflecting the reversal of handedness when the sign of the odd mobility is changed. It is also nonmonotonic in \(|\kappa|\). For fixed \(\tilde\epsilon\), the extrema occur at
\begin{equation}
	|\kappa_\ast|
	=
	\frac{1+\tilde\epsilon}{\tilde\epsilon},
	\qquad
	|\widetilde C_{14}^{\max}|
	=
	\frac{\Theta-1}{4(1+\tilde\epsilon)} .
	\label{eq:ss_kappa_star}
\end{equation}
Therefore weak coupling shifts the optimal oddness to large \(|\kappa|\), while stronger coupling moves the optimum toward \(|\kappa|\simeq1\). At the same time, strong coupling suppresses the maximal dimensionless transverse correlation because the relative motion of the two particles becomes more tightly constrained.

Figure~\ref{fig:odd_correlation} shows the interparticle correlation and its optimal oddness. The odd cross-correlation changes sign with \(\kappa\), reaches extrema at finite \(\pm\kappa_\ast\), and disappears in the even-mobility limit. The odd cross-correlation therefore provides a direct structural signature of the transverse force response, distinct from the irreversible currents already present in scalar two-temperature dimers~\cite{dotsenko2023out}.

\section{Steady-state probability currents}
\label{sec:steady_state_currents}
\vspace{5mm}

The Gaussian steady state derived in Sec.~\ref{sec:steady_state_odd_correlations} fixes the equal-time correlations, but it does not by itself characterize the nonequilibrium dynamics. At stationarity, the continuity equation only requires \(\nabla_{\mathbf R}\cdot\mathbf J_{\mathrm{ss}}=0\); the current \(\mathbf J_{\mathrm{ss}}\) may still circulate in configuration space. This distinction is especially important in the present odd-mobility model. The covariance matrix is determined by the symmetric part of the transport matrix, \(\mathbf D_s\), whereas the full probability current contains the complete matrix \(\mathbf D\), including its antisymmetric part. Thus the odd contribution to transport can be invisible in the Gaussian normalization problem while remaining explicit in the steady current.

In this section, we first express the full configuration-space current in terms of an antisymmetric matrix \(\mathbf N\) that measures the departure from detailed balance. We then decompose \(\mathbf N\) into an even irreversible amplitude and an odd chiral amplitude, before projecting the current onto the one-particle subspaces to obtain the marginal circulating currents.

\subsection{Full configuration-space current}
\label{subsec:full_configuration_current}

For the Gaussian steady state in Eq.~\eqref{eq:ss_gaussian}, one has \(\nabla_{\mathbf R}P_{\mathrm{ss}}=-\mathbf C^{-1}\mathbf R P_{\mathrm{ss}}\). Substitution into the current in Eq.~\eqref{eq:model_current} gives
\begin{equation}
	\mathbf J_{\mathrm{ss}}(\mathbf R)
	=
	\left(
	-\mathbf A+\mathbf D\mathbf C^{-1}
	\right)
	\mathbf R P_{\mathrm{ss}}(\mathbf R).
	\label{eq:current_B_form}
\end{equation}
The steady current is therefore a linear flow field in the configuration vector \(\mathbf R\), weighted by the stationary Gaussian density.

It is useful to rewrite Eq.~\eqref{eq:current_B_form} in a form that isolates the matrix controlling the irreversible circulation. We define
\begin{equation}
	\mathbf N
	=
	\mathbf D-\mathbf A\mathbf C .
	\label{eq:N_definition}
\end{equation}
Then Eq.~\eqref{eq:current_B_form} becomes
\begin{equation}
	\mathbf J_{\mathrm{ss}}(\mathbf R)
	=
	\mathbf N\mathbf C^{-1}\mathbf R P_{\mathrm{ss}}(\mathbf R).
	\label{eq:current_N_form}
\end{equation}
As we show in \ref{app:current_derivations}, the Lyapunov equation in Eq.~\eqref{eq:ss_lyapunov} immediately gives 
\begin{equation}
	\mathbf N+\mathbf N^{\mathrm T}
	=
	\mathbf 0 .
	\label{eq:N_antisymmetric}
\end{equation}
Thus \(\mathbf N\) is antisymmetric. The condition \(\mathbf N=\mathbf 0\) is equivalent to detailed balance in configuration space, since it makes \(\mathbf J_{\mathrm{ss}}\) vanish everywhere. Conversely, \(\mathbf N\neq\mathbf 0\) implies a nonzero solenoidal current circulating around the steady Gaussian density. The structure of \(\mathbf N\) therefore contains the irreversible part of the steady dynamics that cannot be inferred from \(P_{\mathrm{ss}}\) alone.

\subsection{Even and odd current amplitudes}
\label{subsec:even_odd_current_amplitudes}
\begin{figure}[tbp]
	\centering
	\includegraphics[width=1\textwidth]{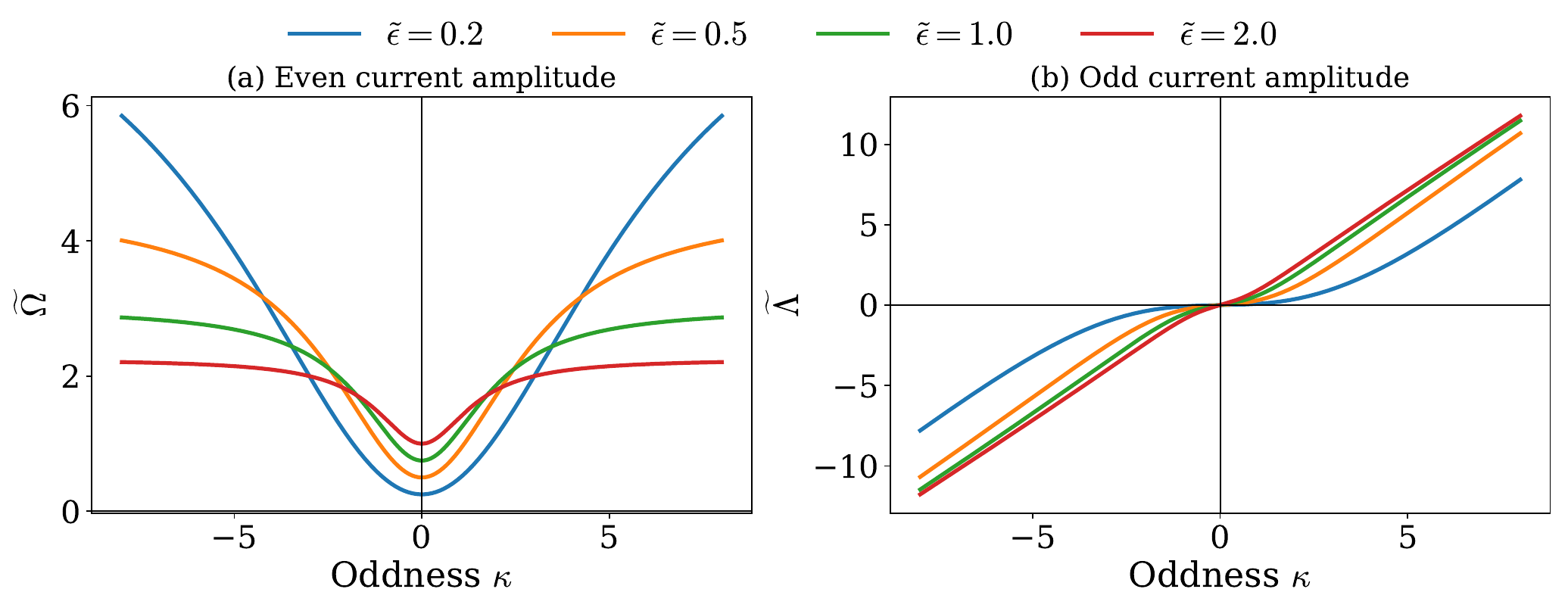}
	\caption{
		Dimensionless even and odd amplitudes of the steady-state probability current for the temperature ratio \(\Theta=T_2/T_1=4.0\). 
		\textbf{(a)} The even current amplitude \(\widetilde{\Omega}=\Omega/D_0\) as a function of the oddness parameter \(\kappa\) for different dimensionless couplings \(\tilde\epsilon\). 
		The amplitude \(\widetilde{\Omega}\) is symmetric under \(\kappa\to-\kappa\) and remains finite at \(\kappa=0\) when \(T_2\neq T_1\) and \(\epsilon\neq0\). 
		\textbf{(b)} The odd current amplitude \(\widetilde{\Lambda}=\Lambda/D_0\), which is antisymmetric under \(\kappa\to-\kappa\) and therefore encodes the handedness of the chiral probability current. 
		Both amplitudes vanish when the reservoirs have equal temperatures or when the particles are uncoupled.
	}
	\label{fig:current_amplitudes}
\end{figure}

For  \(T_2>T_1\) , the antisymmetric matrix \(\mathbf N\) can be decomposed into two scalar amplitudes, given as
\begin{equation}
	\mathbf N
	=
	\begin{pmatrix}
		0 & -\Lambda & \Omega & 0 \\
		\Lambda & 0 & 0 & \Omega \\
		-\Omega & 0 & 0 & \Lambda \\
		0 & -\Omega & -\Lambda & 0
	\end{pmatrix}.
	\label{eq:N_matrix_Omega_Lambda}
\end{equation}

Introducing the reference diffusion coefficient \(D_0=\mu_0T_1\), we write \(\Omega=D_0\widetilde{\Omega}\) and \(\Lambda=D_0\widetilde{\Lambda}\). In terms of \(\tilde\epsilon\), \(\Theta\), and \(\mathcal Q_\kappa\)
the dimensionless amplitudes are
\begin{equation}
	\widetilde{\Omega}
	=
	\frac{
		\tilde\epsilon(1+\tilde\epsilon)(1+\kappa^2)(\Theta-1)
	}
	{
		2\mathcal Q_\kappa
	},
	\qquad
	\widetilde{\Lambda}
	=
	\frac{
		\tilde\epsilon^2\kappa(1+\kappa^2)(\Theta-1)
	}
	{
		2\mathcal Q_\kappa
	} .
	\label{eq:Omega_Lambda_dimensionless}
\end{equation}
The amplitude \(\Omega\) is even under \(\kappa\to-\kappa\), whereas \(\Lambda\) is odd under \(\kappa\to-\kappa\). Thus \(\Omega\) describes the even irreversible part of the configuration-space current, while \(\Lambda\) describes the chiral component generated by odd mobility. Both amplitudes vanish when \(T_1=T_2\) or \(\epsilon=0\), since in either case there is no nonequilibrium exchange channel between the two particles. However, their symmetry with respect to \(\kappa\) is different. The even amplitude \(\widetilde{\Omega}\) is insensitive to the sign of \(\kappa\), while the odd amplitude \(\widetilde{\Lambda}\) changes sign when the handedness of odd transport is reversed. Their ratio is particularly simple and reads
\begin{equation}
	\frac{\widetilde{\Lambda}}{\widetilde{\Omega}}
	=
	\frac{\kappa\tilde\epsilon}{1+\tilde\epsilon}.
	\label{eq:Lambda_Omega_ratio}
\end{equation}
The chiral part of the current therefore becomes more prominent for larger \(|\kappa|\) and stronger coupling.

Figure~\ref{fig:current_amplitudes} illustrates the distinct roles of \(\widetilde{\Omega}\) and \(\widetilde{\Lambda}\). The even component is already present for a coupled two-temperature dimer with \(\kappa=0\), while the odd component requires \(\kappa\neq0\) and changes sign with \(\kappa\). This separation will be useful below when comparing the full configuration-space current with the marginal currents of the individual particles.

\subsection{Marginal densities and circulating currents}
\label{subsec:marginal_currents}
\begin{figure}[tbp]
	\centering
	\includegraphics[width=1\textwidth]{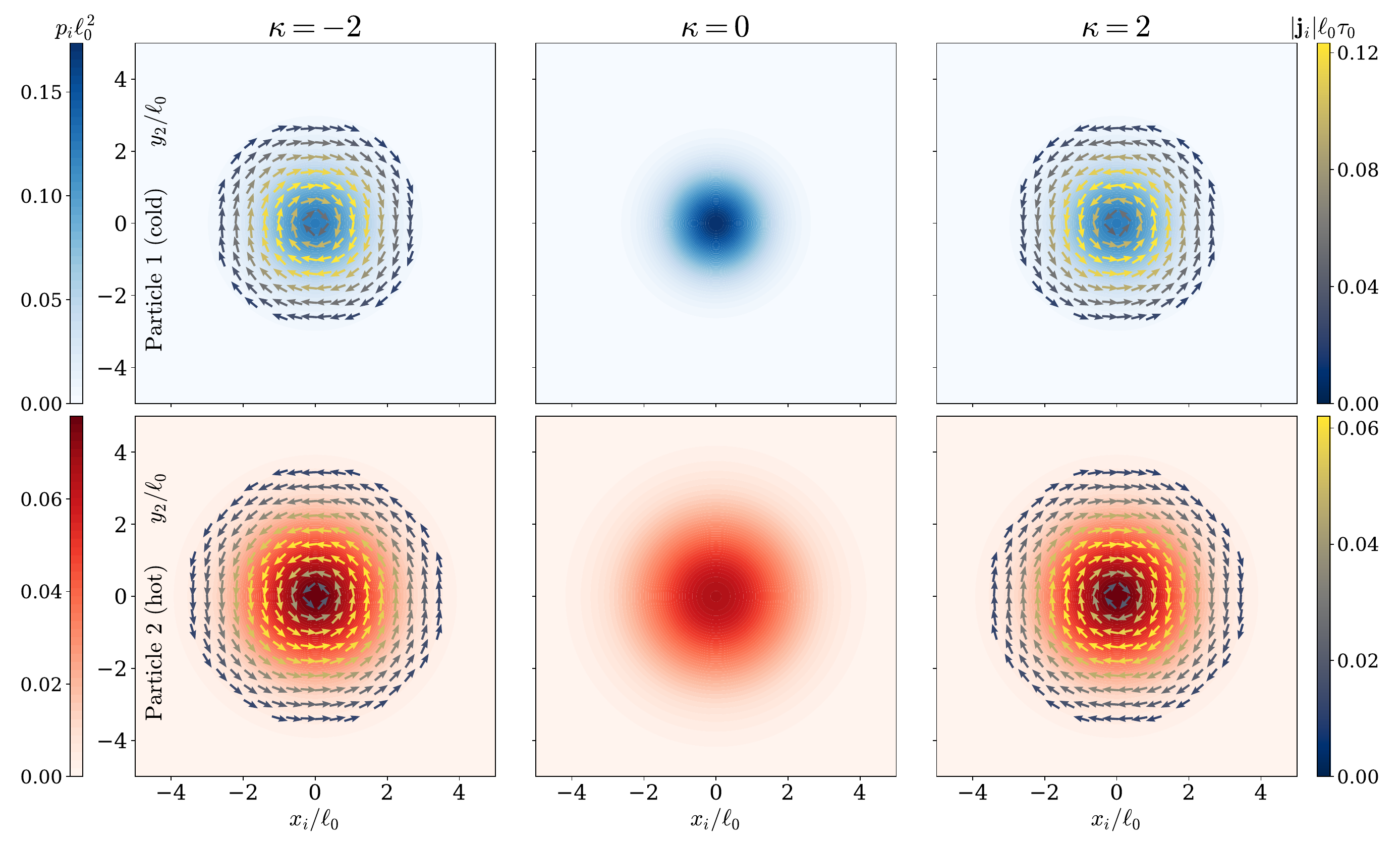}
	\caption{
		Marginal steady-state densities and circulating probability currents for the temperature ratio \(\Theta=T_2/T_1=4.0\). 
		The top row shows the marginal density \(p_1(\mathbf r_1)\) and current \(\mathbf j_1(\mathbf r_1)\) of particle \(1\), while the bottom row shows the corresponding quantities for particle \(2\). 
		The columns correspond to \(\kappa=-2\), \(\kappa=0\), and \(\kappa=2\). 
		Densities are shown in dimensionless form as \(p_i\ell_0^2\), and the color of the arrows represents the dimensionless current magnitude \(|\mathbf j_i|\ell_0\tau_0\), with \(\tau_0=(\mu_0 k)^{-1}\). 
		The arrows are normalized to emphasize the circulation direction; no arrows are shown for \(\kappa=0\), where the marginal currents vanish. 
		For \(T_2>T_1\), particle \(2\) has a broader density distribution and therefore a smaller peak probability density than particle \(1\). 
		Changing the sign of \(\kappa\) reverses the circulation direction, while the two particles rotate in opposite directions for a fixed nonzero \(\kappa\).
	}
	\label{fig:marginal_currents}
\end{figure}

The full current in Eq.~\eqref{eq:current_N_form} is defined in the four-dimensional configuration space of \(\mathbf R\). To obtain the irreversible dynamics visible in the coordinates of each particle separately, we integrate over the coordinates of the other particle. The marginal density and marginal current of particle \(i\) are
\begin{equation}
	p_i(\mathbf r_i)
	=
	\int d\mathbf r_j P_{\mathrm{ss}}(\mathbf r_1,\mathbf r_2),
	\qquad
	\mathbf j_i(\mathbf r_i)
	=
	\int d\mathbf r_j \mathbf J_i(\mathbf r_1,\mathbf r_2),
	\qquad
	j\neq i ,
	\label{eq:marginal_definitions}
\end{equation}
where \(\mathbf J_i\) denotes the two-dimensional part of the full current associated with particle \(i\).

Since the diagonal blocks of the covariance matrix are isotropic, the one-particle marginal densities are radial Gaussians, which can be derived and written as (see \ref{app:current_derivations} for details)
\begin{equation}
	p_1(\mathbf r_1)
	=
	\frac{1}{2\pi C_{11}}
	\exp\left[
	-\frac{|\mathbf r_1|^2}{2C_{11}}
	\right],
	\qquad
	p_2(\mathbf r_2)
	=
	\frac{1}{2\pi C_{33}}
	\exp\left[
	-\frac{|\mathbf r_2|^2}{2C_{33}}
	\right].
	\label{eq:marginal_densities}
\end{equation}
The variances \(C_{11}\) and \(C_{33}\) are given in Sec.~\ref{sec:steady_state_odd_correlations}, Eqs.~\eqref{eq:ss_sigma1} and \eqref{eq:ss_sigma2}. Equation~\eqref{eq:ss_width_difference} implies that for \(T_2>T_1\), \(C_{33}>C_{11}\). Consequently, the hotter particle has a wider marginal distribution and, as a result, a reduced peak density.


The corresponding marginal currents take the purely circulating form
\begin{equation}
	\mathbf j_i(\mathbf r_i)
	=
	\omega_i\boldsymbol{\varepsilon}\mathbf r_i p_i(\mathbf r_i),
	\label{eq:marginal_angular_amplitudes}
\end{equation}
where \(\omega_1=-\frac{\Lambda}{C_{11}}\) and \(\omega_2=\frac{\Lambda}{C_{33}}\). Because \(\boldsymbol{\varepsilon}\mathbf r_i\) is perpendicular to \(\mathbf r_i\), the marginal currents are tangential
\begin{equation}
	\mathbf r_i\cdot\mathbf j_i(\mathbf r_i)=0 .
	\label{eq:marginal_current_tangential}
\end{equation}
The marginal current therefore describes a local circulation of the probability density, with \(\omega_i\) setting the angular current amplitude. This form is not an additional assumption: it follows from integrating the exact four-dimensional steady current over the coordinates of the other particle as we show in \ref{app:marginal_currents}.

For \(T_2>T_1\) and \(\kappa>0\), the odd amplitude satisfies \(\Lambda>0\), and therefore \(\omega_1<0\) while \(\omega_2>0\). The two one-particle currents are thus counter-rotating. Reversing \(\kappa\) reverses both circulation directions, whereas the marginal currents vanish for \(\kappa=0\), \(\epsilon=0\), or \(T_1=T_2\). Thus the scalar two-temperature dimer can possess a nonzero full configuration-space current, but its one-particle marginal currents vanish unless the odd chiral amplitude \(\Lambda\) is present.

Figure~\ref{fig:marginal_currents} shows that the one-particle steady states are not characterized by static density profiles alone. Although each marginal density is radially symmetric, the corresponding marginal current circulates around the trap center when \(\kappa\neq0\). These circulating currents provide a direct real-space manifestation of the odd part of the configuration-space current encoded by \(\widetilde{\Lambda}\).

\section{Heat transfer and entropy production}
\label{sec:heat_entropy}
\vspace{5mm}
\begin{figure}[tbp]
	\centering
	\includegraphics[width=1\textwidth]{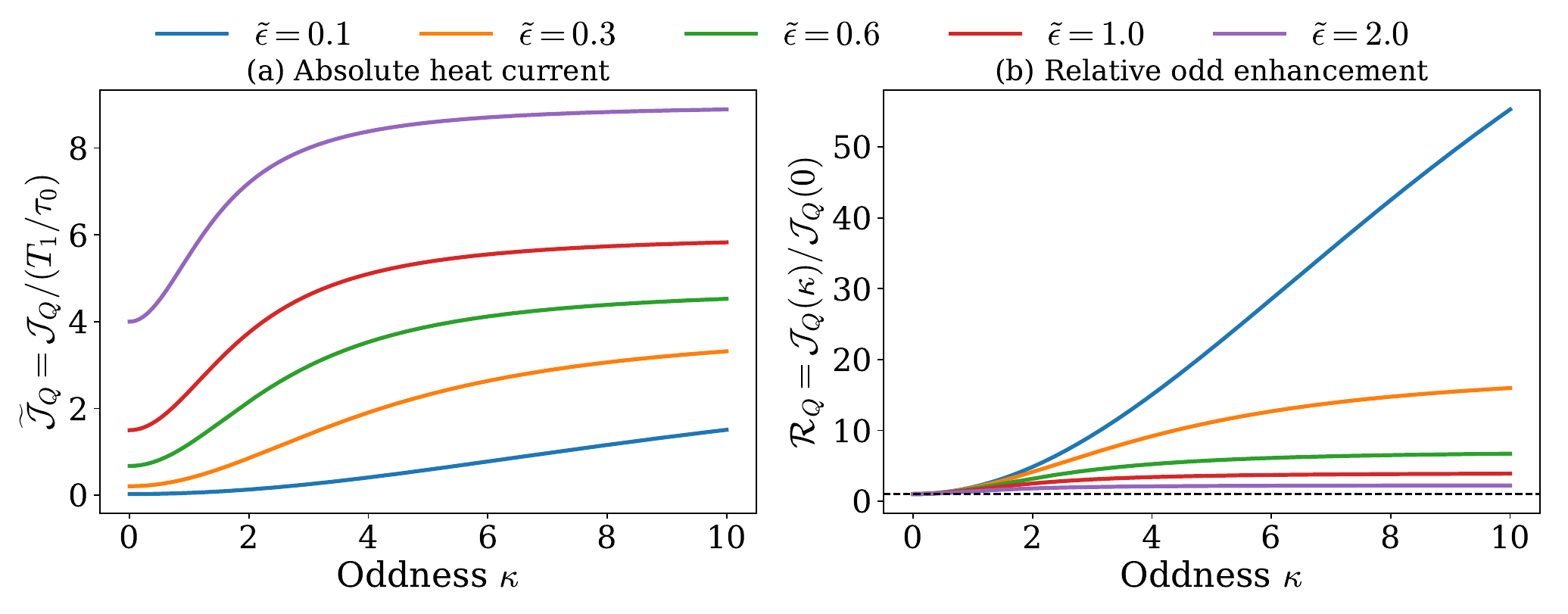}
	\caption{
		Dimensionless heat transfer and odd enhancement for the temperature ratio \(\Theta=4.0\). \textbf{(a)} Absolute heat current from the hot reservoir to the cold reservoir, \(\widetilde{\mathcal J}_Q=\mathcal J_Q/(T_1/\tau_0)\), as a function of the oddness parameter \(\kappa\) for different dimensionless couplings \(\tilde\epsilon\). The absolute heat current increases with both \(\tilde\epsilon\) and \(|\kappa|\), but it vanishes as \(\tilde\epsilon\to0\) because the two particles are then uncoupled. \textbf{(b)} Relative odd enhancement \(\mathcal R_Q=\mathcal J_Q(\kappa)/\mathcal J_Q(0)\). The ratio can become large at weak coupling, although the corresponding absolute heat current remains small. The scalar heat current is even in \(\kappa\): reversing \(\kappa\) reverses the chirality of the probability currents but does not change the net heat-transfer rate.
	}
	\label{fig:heat_transfer}
\end{figure}
The steady currents derived above characterize irreversible motion in configuration space. They also determine the thermodynamic exchange with the two reservoirs, because energy is stored only in the conservative potential \(U\). At steady state, the mean potential energy is constant, so any heat injected by one reservoir must be extracted by the other. This section uses the exact probability currents to obtain the heat-transfer rate and the associated entropy production, and to clarify how odd mobility affects scalar thermodynamic observables.

\subsection{Heat currents}
\label{subsec:heat_currents}

We define \(\dot Q_i\) as the heat current injected by reservoir \(i\) into the configurational degrees of freedom
\begin{equation}
	\dot Q_i
	=
	\int d\mathbf R \,
	\mathbf J_{i,\mathrm{ss}}(\mathbf R)
	\cdot
	\nabla_i U(\mathbf r_1,\mathbf r_2).
	\label{eq:heat_current_definition}
\end{equation}
Here \(\mathbf J_{i,\mathrm{ss}}\) is the two-dimensional component of the full steady-state current associated with particle \(i\). With this convention, \(\dot Q_i>0\) means that reservoir \(i\) injects heat into the system. Since \(\langle U\rangle_{\mathrm{ss}}\) is time independent, energy conservation gives
\begin{equation}
	\dot Q_1+\dot Q_2=0 .
	\label{eq:heat_balance}
\end{equation}
For \(T_2>T_1\), heat is transferred from the reservoir coupled to particle \(2\) to the reservoir coupled to particle \(1\). We therefore define the positive heat-transfer rate
\begin{equation}
	\mathcal J_Q
	=
	\dot Q_2
	=
	-\dot Q_1 .
	\label{eq:positive_heat_current}
\end{equation}
Using the exact Gaussian steady state, one obtains (see ~\ref{app:heat_current_derivation} for details)
\begin{equation}
	\widetilde{\mathcal J}_Q
	=
	(\Theta-1)
	\frac{
		\tilde\epsilon^2(1+\tilde\epsilon)(1+\kappa^2)
	}
	{
		\mathcal Q_\kappa
	} .
	\label{eq:dimensionless_heat_current}
\end{equation}
where  \(\widetilde{\mathcal J}_Q=\mathcal J_Q/(T_1/\tau_0)\) with \(\tau_0=(\mu_0 k)^{-1}\).

 Equation~\eqref{eq:dimensionless_heat_current} shows that heat transfer requires both a temperature difference and a coupling between the particles. It vanishes for \(\Theta=1\), where the reservoirs are at equilibrium, and for \(\tilde\epsilon=0\), where the two particles are dynamically disconnected. The odd mobility modifies the conductance of this elastic heat-transfer channel through the factor \((1+\kappa^2)/\mathcal Q_\kappa\), but it does not change the direction of heat flow, which is fixed by the sign of \(T_2-T_1\).

\subsection{Interaction-controlled heat transfer and odd enhancement}
\label{subsec:heat_odd_enhancement}

A useful way to isolate the role of odd mobility is to compare the heat current at finite \(\kappa\) with that of the corresponding scalar-mobility dimer. We define
\begin{equation}
	\mathcal R_Q
	=
	\frac{
		\mathcal J_Q(\kappa)
	}
	{
		\mathcal J_Q(0)
	}
	=
	\frac{
		(1+\kappa^2)(1+\tilde\epsilon)^2
	}
	{
		(1+\tilde\epsilon)^2
		+
		\kappa^2\tilde\epsilon^2 
	} .
	\label{eq:heat_current_ratio}
\end{equation}
This ratio satisfies \(\mathcal R_Q\geq1\), with equality only at \(\kappa=0\). In the weak-coupling regime, \(\mathcal R_Q\to1+\kappa^2\), so the relative odd enhancement can be large. However, the absolute current in Eq.~\eqref{eq:dimensionless_heat_current} remains small in the same limit because \(\widetilde{\mathcal J}_Q\propto\tilde\epsilon^2\). Thus weak coupling can strongly enhance a heat current that is itself small. In the opposite regime of strong coupling, the relative enhancement is reduced because the two particles are tightly constrained by the spring.

Figure~\ref{fig:heat_transfer} illustrates the distinction between absolute heat transfer and relative enhancement. The absolute heat current is controlled primarily by the strength of the elastic exchange channel, while \(\mathcal R_Q\) measures how much odd mobility increases the heat current relative to the even-mobility case. Unlike the chiral current amplitude \(\widetilde{\Lambda}\), the heat current is a scalar quantity and is therefore even under \(\kappa\to-\kappa\). Changing the sign of \(\kappa\) reverses the circulation direction of the probability currents but leaves \(\mathcal J_Q\) unchanged.

\subsection{Entropy production}
\label{subsec:entropy_production}

The entropy production rate follows directly from the reservoir heat currents. With the sign convention of Eq.~\eqref{eq:heat_current_definition},
\begin{equation}
	\dot S_{\mathrm{tot}}
	=
	-\sum_{i=1}^{2}
	\frac{\dot Q_i}{T_i}.
	\label{eq:entropy_production_definition}
\end{equation}
Using \(\dot Q_2=\mathcal J_Q\) and \(\dot Q_1=-\mathcal J_Q\), this becomes
\begin{equation}
	\tau_0\dot S_{\mathrm{tot}}
=
\widetilde{\mathcal J}_Q
\left(
1-\frac{1}{\Theta}
\right).
	\label{eq:entropy_production_heat_current}
\end{equation}
For \(\Theta>1\), Eq.~\eqref{eq:entropy_production_heat_current} is non-negative because \(\mathcal J_Q>0\). 
Thus entropy production is governed by the same interaction-controlled heat-transfer channel as \(\mathcal J_Q\). It vanishes at equilibrium, \(\Theta=1\), and in the uncoupled limit, \(\tilde\epsilon=0\). Like the heat current, it is even under \(\kappa\to-\kappa\): odd mobility changes the handedness of currents and correlations, but scalar dissipation is insensitive to the sign of the odd response.

\section{Discussion and conclusions}
\vspace{5mm}

We have studied an exactly solvable two-temperature Brownian dimer with odd mobility. The system consists of two overdamped particles in a two-dimensional plane, each confined by an isotropic harmonic trap, coupled by an elastic spring, and connected to a thermal reservoir. The conservative potential is harmonic and rotationally invariant, while the mobility contains an antisymmetric contribution that converts conservative forces into transverse drift. This makes the model a minimal setting in which heat exchange, irreversible probability currents, and odd transverse response can be analyzed without imposing any external torque, trap anisotropy, or nonconservative force.

We derived an exact steady-state probability density which is Gaussian. Its one-particle marginals remain isotropic, but the full covariance matrix acquires a transverse interparticle correlation between orthogonal coordinates of the two particles. This correlation is absent if any of the three essential ingredients is removed: temperature imbalance, elastic coupling, or odd mobility. It also reverses sign when the handedness of the odd response is reversed and reaches its largest magnitude at finite oddness. Thus odd mobility does not simply broaden or deform the individual particle distributions; it produces a handed structural correlation between the two particles.

The steady probability currents reveal the dynamical counterpart of this structural signature. The stationary density alone does not show the full nonequilibrium character of the system, because a Gaussian density can coexist with circulating currents in configuration space. The full current separates into a part already present in scalar two-temperature dimers and a genuinely chiral part generated by odd mobility. After marginalization, this chiral contribution appears as circulating one-particle currents inside the traps, with the hot and cold particles rotating in opposite directions. These marginal circulations vanish in the scalar-mobility limit, even though irreversible currents may still persist in the full configuration space.

We also obtained exact expressions for the heat currents and entropy production from the steady-state probability fluxes. We found that, in the effective odd-mobility model considered here, odd transport enhances the thermal conductance between the two reservoirs. At the same time, the heat current and entropy production are scalar observables: they are insensitive to the sign of the odd response. Reversing the handedness therefore reverses the transverse correlations and odd flux components, but not the direction of net heat transfer or the total dissipation. This separation between handed structural and dynamical observables on the one hand, and scalar thermodynamic observables on the other, is the main physical result of the dimer.

The present model provides a controlled reference point for more complex odd two-temperature systems. A direct extension is to couple Brownian-gyrator units instead of isotropic trapped particles, thereby combining internal anisotropic heat-engine-like dynamics with interparticle heat exchange and odd mobility~\cite{filliger2007brownian, abdoli2026quadrupolar}. A second direction is to study many-particle hot--cold mixtures with odd transport, where the transverse pair correlations identified here may become elementary building blocks of collective chiral organization, rotating clusters, or modified demixing~\cite{weber2016binary,ilker2020phase,mccarthy2024demixing}. Another analytically accessible extension would be to consider two traps with different stiffnesses, which would test how trap asymmetry modifies the transverse correlations, marginal circulations, and heat transfer while preserving the exact solvability of the linear model.
Finally, comparing the effective odd-mobility description with explicitly underdamped Lorentz-force dynamics would clarify how microscopic gyrotropic motion maps onto overdamped odd transport in heat-conducting Brownian matter.

A possible experimental realization could build on optically trapped colloidal dimers with reservoir temperatures controlled by external noise applied to the traps~\cite{berut2014energy,berut2016stationary}. In such a setup, the elastic coupling and temperature imbalance are already directly tunable. The odd transport element could be introduced by embedding the trapped particles in a bath of circle swimmers or other rotating active particles, whose persistent handed motion generates an effective transverse response at coarse-grained scales~\cite{goerlich2026particle}. Alternatively, feedback-controlled optical forces could be used to impose a force component perpendicular to the conservative restoring force~\cite{lichtner2010feedback, gernert2015enhancement, gernert2016feedback}. Such realizations would allow one to test the predicted transverse correlations, counter-rotating marginal currents, and odd enhancement of heat transfer in a controlled Brownian system.

\ack{This work was funded by the Deutsche Forschungsgemeinschaft (DFG, German Research Foundation) under project number 556762905 — AB 1083/1-1.}



\data{All data that support the findings of this study are included within the article (and any supplementary files).}



\appendix

\renewcommand{\thesection}{Appendix~\Alph{section}}

\numberwithin{equation}{section}
\renewcommand{\theequation}{\Alph{section}\arabic{equation}}

\section{Exact Gaussian steady state}
\label{app:steady_state_derivation}
\vspace{5mm}
This appendix gives the derivation of the exact steady-state covariance used in Sec.~\ref{sec:steady_state_odd_correlations}. We use the coordinate vector \(\mathbf R=(x_1,y_1,x_2,y_2)^{\mathrm T}\), for which the harmonic potential in Eq.~\eqref{eq:model_potential} can be written as
\begin{equation}
	U(\mathbf R)
	=
	\frac{1}{2}\mathbf R^{\mathrm T}\mathbf K\mathbf R ,
	\label{eq:app_quadratic_potential}
\end{equation}
with the stiffness matrix
\begin{equation}
	\mathbf K
	=
	\left(
	\begin{array}{cc}
		(k+\epsilon)\mathbf I & -\epsilon\mathbf I \\
		-\epsilon\mathbf I & (k+\epsilon)\mathbf I
	\end{array}
	\right).
	\label{eq:app_K_block}
\end{equation}
The corresponding four-dimensional mobility matrix is
\begin{equation}
	\boldsymbol{\mathcal M}
	=
	\left(
	\begin{array}{cc}
		\boldsymbol{\mu} & \mathbf 0 \\
		\mathbf 0 & \boldsymbol{\mu}
	\end{array}
	\right),
	\qquad
	\boldsymbol{\mu}
	=
	\mu_0
	\left(
	\mathbf I+\kappa\boldsymbol{\varepsilon}
	\right).
	\label{eq:app_M_block}
\end{equation}
With the convention \(\varepsilon_{xy}=1\) and \(\varepsilon_{yx}=-1\), this becomes, in the coordinate ordering \((x_1,y_1,x_2,y_2)\),
\begin{equation}
	\boldsymbol{\mathcal M}
	=
	\mu_0
	\begin{pmatrix}
		1 & \kappa & 0 & 0 \\
		-\kappa & 1 & 0 & 0 \\
		0 & 0 & 1 & \kappa \\
		0 & 0 & -\kappa & 1
	\end{pmatrix}.
	\label{eq:app_M_explicit}
\end{equation}

The drift matrix appearing in the Fokker--Planck current is
\begin{equation}
	\mathbf A
	=
	\boldsymbol{\mathcal M}\mathbf K .
	\label{eq:app_A_definition}
\end{equation}
Explicitly,
\begin{equation}
	\mathbf A
	=
	\mu_0
	\begin{pmatrix}
		k+\epsilon & \kappa(k+\epsilon) & -\epsilon & -\kappa\epsilon \\
		-\kappa(k+\epsilon) & k+\epsilon & \kappa\epsilon & -\epsilon \\
		-\epsilon & -\kappa\epsilon & k+\epsilon & \kappa(k+\epsilon) \\
		\kappa\epsilon & -\epsilon & -\kappa(k+\epsilon) & k+\epsilon
	\end{pmatrix}.
	\label{eq:app_A_explicit}
\end{equation}
The temperature-weighted transport matrix is
\begin{equation}
	\mathbf D
	=
	\left(
	\begin{array}{cc}
		T_1\boldsymbol{\mu} & \mathbf 0 \\
		\mathbf 0 & T_2\boldsymbol{\mu}
	\end{array}
	\right),
	\label{eq:app_D_block}
\end{equation}
or, equivalently,
\begin{equation}
	\mathbf D
	=
	\mu_0
	\begin{pmatrix}
		T_1 & \kappa T_1 & 0 & 0 \\
		-\kappa T_1 & T_1 & 0 & 0 \\
		0 & 0 & T_2 & \kappa T_2 \\
		0 & 0 & -\kappa T_2 & T_2
	\end{pmatrix}.
	\label{eq:app_D_explicit}
\end{equation}
Only the symmetric part of \(\mathbf D\) enters the covariance equation. It is
\begin{equation}
	\mathbf D_s
	=
	\frac{1}{2}
	\left(
	\mathbf D+\mathbf D^{\mathrm T}
	\right)
	=
	\mu_0\,
	\mathrm{diag}
	\left(
	T_1,T_1,T_2,T_2
	\right).
	\label{eq:app_Ds}
\end{equation}
The antisymmetric part of \(\mathbf D\) contributes to the probability current, but it drops out of the second-order diffusion operator because \(\partial_\alpha\partial_\beta P\) is symmetric under \(\alpha\leftrightarrow\beta\).

The Fokker--Planck equation can therefore be written as
\begin{equation}
	\partial_tP
	=
	-\nabla_{\mathbf R}\cdot\mathbf J,
	\qquad
	\mathbf J
	=
	-\mathbf A\mathbf R\,P
	-
	\mathbf D\nabla_{\mathbf R}P .
	\label{eq:app_fpe_matrix}
\end{equation}
Since \(\mathbf A\) and \(\mathbf D\) are constant and the drift is linear in \(\mathbf R\), the steady state is Gaussian,
\begin{equation}
	P_{\mathrm{ss}}(\mathbf R)
	=
	\frac{1}{(2\pi)^2\sqrt{\det\mathbf C}}
	\exp\left[
	-\frac{1}{2}
	\mathbf R^{\mathrm T}\mathbf C^{-1}\mathbf R
	\right],
	\label{eq:app_gaussian}
\end{equation}
where \(\mathbf C=\langle\mathbf R\mathbf R^{\mathrm T}\rangle_{\mathrm{ss}}\). The covariance follows from the evolution equation for the second moment. Multiplying Eq.~\eqref{eq:app_fpe_matrix} by \(\mathbf R\mathbf R^{\mathrm T}\), integrating over \(\mathbf R\), and using integration by parts gives
\begin{equation}
	\frac{d\mathbf C}{dt}
	=
	-\mathbf A\mathbf C
	-
	\mathbf C\mathbf A^{\mathrm T}
	+
	2\mathbf D_s .
	\label{eq:app_covariance_dynamics}
\end{equation}
At stationarity, \(d\mathbf C/dt=\mathbf 0\), and hence
\begin{equation}
	\mathbf A\mathbf C
	+
	\mathbf C\mathbf A^{\mathrm T}
	=
	2\mathbf D_s .
	\label{eq:app_lyapunov}
\end{equation}
This is the Lyapunov equation quoted in Eq.~\eqref{eq:ss_lyapunov}.

Solving Eq.~\eqref{eq:app_lyapunov} gives the covariance matrix
\begin{equation}
	\mathbf C
	=
	\begin{pmatrix}
		C_{11} & 0 & C_{13} & C_{14} \\
		0 & C_{11} & C_{23} & C_{13} \\
		C_{13} & C_{23} & C_{33} & 0 \\
		C_{14} & C_{13} & 0 & C_{33}
	\end{pmatrix}.
	\label{eq:app_covariance_structure}
\end{equation}
It is useful to introduce
\begin{equation}
	\Delta_\kappa
	=
	k^2+2k\epsilon+(1+\kappa^2)\epsilon^2
	=
	(k+\epsilon)^2+\kappa^2\epsilon^2 .
	\label{eq:app_delta_kappa}
\end{equation}
The diagonal entries are
\begin{equation}
	C_{11}
	=
	\frac{
		(k+\epsilon)
		\left[
		2k^2T_1
		+
		4k\epsilon T_1
		+
		(1+\kappa^2)\epsilon^2(T_1+T_2)
		\right]
	}
	{
		2k(k+2\epsilon)\Delta_\kappa
	},
	\label{eq:app_C11}
\end{equation}
and
\begin{equation}
	C_{33}
	=
	\frac{
		(k+\epsilon)
		\left[
		2k^2T_2
		+
		4k\epsilon T_2
		+
		(1+\kappa^2)\epsilon^2(T_1+T_2)
		\right]
	}
	{
		2k(k+2\epsilon)\Delta_\kappa
	}.
	\label{eq:app_C33}
\end{equation}
The ordinary interparticle correlations are
\begin{equation}
	C_{13}
	=
	\langle x_1x_2\rangle
	=
	\langle y_1y_2\rangle
	=
	\frac{\epsilon(T_1+T_2)}
	{2k(k+2\epsilon)} .
	\label{eq:app_C13}
\end{equation}
The odd transverse interparticle correlations are
\begin{equation}
	C_{14}
	=
	\langle x_1y_2\rangle
	=
	\frac{
		\epsilon\kappa(T_2-T_1)
	}
	{
		2\Delta_\kappa
	},
	\qquad
	C_{23}
	=
	\langle y_1x_2\rangle
	=
	-C_{14}.
	\label{eq:app_C14_C23}
\end{equation}
For \(T_2>T_1\), the sign of \(C_{14}\) is set by the sign of \(\kappa\), while \(C_{23}\) has the opposite sign.

The determinant required for the normalization of Eq.~\eqref{eq:app_gaussian} is
\begin{equation}
	\det\mathbf C
	=
	\frac{
		\left[
		4k^2T_1T_2
		+
		8k\epsilon T_1T_2
		+
		(1+\kappa^2)\epsilon^2(T_1+T_2)^2
		\right]^2
	}
	{
		16k^2(k+2\epsilon)^2\Delta_\kappa^2
	} .
	\label{eq:app_detC}
\end{equation}

Finally, we derive the dimensionless form of the odd correlation. Using \(\ell_0^2=T_1/k\), \(\tilde\epsilon=\epsilon/k\), and \(\Theta=T_2/T_1\), Eq.~\eqref{eq:app_C14_C23} gives
\begin{equation}
	\widetilde C_{14}
	=
	\frac{C_{14}}{\ell_0^2}
	=
	\frac{
		\tilde\epsilon\,\kappa(\Theta-1)
	}
	{
		2\left[
		(1+\tilde\epsilon)^2
		+
		\kappa^2\tilde\epsilon^2
		\right]
	} .
	\label{eq:app_C14_dimensionless}
\end{equation}
The extrema of \(\widetilde C_{14}\) with respect to \(\kappa\) follow from \(\partial\widetilde C_{14}/\partial\kappa=0\), giving
\begin{equation}
	|\kappa_\ast|
	=
	\frac{1+\tilde\epsilon}{\tilde\epsilon},
	\qquad
	|\widetilde C_{14}^{\max}|
	=
	\frac{\Theta-1}{4(1+\tilde\epsilon)} .
	\label{eq:app_C14_extrema}
\end{equation}
Thus the odd transverse correlation is optimized at finite \(|\kappa|\), rather than increasing monotonically with the oddness.

\section{Steady-state probability currents}
\label{app:current_derivations}
\vspace{5mm}
This appendix gives the algebra leading to the configuration-space current, the current amplitudes \(\Omega\) and \(\Lambda\), and the marginal one-particle currents used in Sec.~\ref{sec:steady_state_currents}.

\subsection{Configuration-space current}
\label{app:configuration_current}

The steady-state probability current follows by inserting the Gaussian steady state in Eq.~\eqref{eq:ss_gaussian} into the current in Eq.~\eqref{eq:model_current}. Since
\begin{equation}
	\nabla_{\mathbf R}P_{\mathrm{ss}}
	=
	-\mathbf C^{-1}\mathbf R\,P_{\mathrm{ss}},
	\label{eq:app_grad_gaussian}
\end{equation}
one obtains
\begin{equation}
	\mathbf J_{\mathrm{ss}}(\mathbf R)
	=
	\left(
	-\mathbf A+\mathbf D\mathbf C^{-1}
	\right)
	\mathbf R\,P_{\mathrm{ss}}(\mathbf R).
	\label{eq:app_current_B_form}
\end{equation}
Equivalently, defining
\begin{equation}
	\mathbf N
	=
	\mathbf D-\mathbf A\mathbf C ,
	\label{eq:app_N_definition}
\end{equation}
the current can be written as
\begin{equation}
	\mathbf J_{\mathrm{ss}}(\mathbf R)
	=
	\mathbf N\mathbf C^{-1}\mathbf R\,P_{\mathrm{ss}}(\mathbf R).
	\label{eq:app_current_N_form}
\end{equation}
The matrix \(\mathbf N\) measures the part of the transport that is not balanced by the stationary covariance. Using the Lyapunov equation,
\begin{equation}
	\mathbf A\mathbf C+\mathbf C\mathbf A^{\mathrm T}
	=
	2\mathbf D_s
	=
	\mathbf D+\mathbf D^{\mathrm T},
	\label{eq:app_current_lyapunov}
\end{equation}
we find
\begin{equation}
	\mathbf N+\mathbf N^{\mathrm T}
	=
	\mathbf D+\mathbf D^{\mathrm T}
	-
	\mathbf A\mathbf C
	-
	\mathbf C\mathbf A^{\mathrm T}
	=
	\mathbf 0 .
	\label{eq:app_N_antisymmetric}
\end{equation}
Thus \(\mathbf N\) is antisymmetric. The condition \(\mathbf N=\mathbf 0\) is equivalent to a vanishing stationary current and hence to detailed balance in configuration space.

For \(T_2>T_1\), the explicit form of \(\mathbf N\) is
\begin{equation}
	\mathbf N
	=
	\begin{pmatrix}
		0 & -\Lambda & \Omega & 0 \\
		\Lambda & 0 & 0 & \Omega \\
		-\Omega & 0 & 0 & \Lambda \\
		0 & -\Omega & -\Lambda & 0
	\end{pmatrix}.
	\label{eq:app_N_matrix}
\end{equation}
Here,
\begin{equation}
	\Omega
	=
	\frac{
		\mu_0\epsilon(k+\epsilon)(1+\kappa^2)(T_2-T_1)
	}
	{
		2\Delta_\kappa
	},
	\qquad
	\Lambda
	=
	\frac{
		\mu_0\epsilon^2\kappa(1+\kappa^2)(T_2-T_1)
	}
	{
		2\Delta_\kappa
	},
	\label{eq:app_Omega_Lambda_dimensional}
\end{equation}
with \(\Delta_\kappa=(k+\epsilon)^2+\kappa^2\epsilon^2\). The amplitude \(\Omega\) is even in \(\kappa\), whereas \(\Lambda\) is odd in \(\kappa\). This is the origin of the separation between an even irreversible component and a chiral odd component of the stationary current.

\subsection{Marginal densities and marginal currents}
\label{app:marginal_currents}

The marginal density of particle \(i\) is obtained by integrating \(P_{\mathrm{ss}}\) over the coordinates of the other particle. Since the covariance matrix in Eq.~\eqref{eq:ss_covariance_structure} has isotropic diagonal blocks, the marginal densities are
\begin{equation}
	p_1(\mathbf r_1)
	=
	\frac{1}{2\pi C_{11}}
	\exp\left[
	-\frac{|\mathbf r_1|^2}{2C_{11}}
	\right],
	\qquad
	p_2(\mathbf r_2)
	=
	\frac{1}{2\pi C_{33}}
	\exp\left[
	-\frac{|\mathbf r_2|^2}{2C_{33}}
	\right].
	\label{eq:app_marginal_densities}
\end{equation}
The corresponding marginal currents are defined by
\begin{equation}
	\mathbf j_1(\mathbf r_1)
	=
	\int d\mathbf r_2\,\mathbf J_1(\mathbf r_1,\mathbf r_2),
	\qquad
	\mathbf j_2(\mathbf r_2)
	=
	\int d\mathbf r_1\,\mathbf J_2(\mathbf r_1,\mathbf r_2),
	\label{eq:app_marginal_current_definitions}
\end{equation}
where \(\mathbf J_1\) and \(\mathbf J_2\) denote the first and last two components of \(\mathbf J_{\mathrm{ss}}\), respectively.

For particle \(1\), the current can be written as
\begin{equation}
	\mathbf j_1(\mathbf r_1)
	=
	-\boldsymbol{\mu}
	\left[
	p_1(\mathbf r_1)
	\left\langle
	\nabla_1U
	\mid
	\mathbf r_1
	\right\rangle
	+
	T_1\nabla_1p_1(\mathbf r_1)
	\right].
	\label{eq:app_j1_conditional}
\end{equation}
The conditional mean of \(\mathbf r_2\) at fixed \(\mathbf r_1\) is
\begin{equation}
	\left\langle
	\mathbf r_2
	\mid
	\mathbf r_1
	\right\rangle
	=
	\frac{1}{C_{11}}
	\begin{pmatrix}
		C_{13} & C_{23} \\
		C_{14} & C_{13}
	\end{pmatrix}
	\mathbf r_1 .
	\label{eq:app_conditional_r2_given_r1}
\end{equation}
Using
\begin{equation}
	\left\langle
	\nabla_1U
	\mid
	\mathbf r_1
	\right\rangle
	=
	(k+\epsilon)\mathbf r_1
	-
	\epsilon
	\left\langle
	\mathbf r_2
	\mid
	\mathbf r_1
	\right\rangle,
	\qquad
	\nabla_1p_1
	=
	-\frac{\mathbf r_1}{C_{11}}p_1,
	\label{eq:app_j1_ingredients}
\end{equation}
one obtains after simplification
\begin{equation}
	\mathbf j_1(\mathbf r_1)
	=
	\omega_1\boldsymbol{\varepsilon}\mathbf r_1\,p_1(\mathbf r_1),
	\label{eq:app_j1_circulating}
\end{equation}
where
\begin{equation}
	\omega_1
	=
	-
	\frac{
		\mu_0\kappa\epsilon^2 k(k+2\epsilon)(1+\kappa^2)(T_2-T_1)
	}
	{
		(k+\epsilon)
		\left[
		2k^2T_1
		+
		4k\epsilon T_1
		+
		(1+\kappa^2)\epsilon^2(T_1+T_2)
		\right]
	} .
	\label{eq:app_omega1}
\end{equation}

Similarly, for particle \(2\),
\begin{equation}
	\mathbf j_2(\mathbf r_2)
	=
	-\boldsymbol{\mu}
	\left[
	p_2(\mathbf r_2)
	\left\langle
	\nabla_2U
	\mid
	\mathbf r_2
	\right\rangle
	+
	T_2\nabla_2p_2(\mathbf r_2)
	\right],
	\label{eq:app_j2_conditional}
\end{equation}
with
\begin{equation}
	\left\langle
	\mathbf r_1
	\mid
	\mathbf r_2
	\right\rangle
	=
	\frac{1}{C_{33}}
	\begin{pmatrix}
		C_{13} & C_{14} \\
		C_{23} & C_{13}
	\end{pmatrix}
	\mathbf r_2 .
	\label{eq:app_conditional_r1_given_r2}
\end{equation}
This gives
\begin{equation}
	\mathbf j_2(\mathbf r_2)
	=
	\omega_2\boldsymbol{\varepsilon}\mathbf r_2\,p_2(\mathbf r_2),
	\label{eq:app_j2_circulating}
\end{equation}
where
\begin{equation}
	\omega_2
	=
	\frac{
		\mu_0\kappa\epsilon^2 k(k+2\epsilon)(1+\kappa^2)(T_2-T_1)
	}
	{
		(k+\epsilon)
		\left[
		2k^2T_2
		+
		4k\epsilon T_2
		+
		(1+\kappa^2)\epsilon^2(T_1+T_2)
		\right]
	} .
	\label{eq:app_omega2}
\end{equation}
For \(T_2>T_1\) and \(\kappa>0\), Eqs.~\eqref{eq:app_omega1} and \eqref{eq:app_omega2} imply \(\omega_1<0\) and \(\omega_2>0\). Thus the two marginal currents circulate in opposite directions. Since \(\boldsymbol{\varepsilon}\mathbf r_i\) is perpendicular to \(\mathbf r_i\), both marginal currents are tangential:
\begin{equation}
	\mathbf r_i\cdot\mathbf j_i(\mathbf r_i)
	=
	0,
	\qquad
	i=1,2 .
	\label{eq:app_marginal_tangential}
\end{equation}
The marginal currents vanish when \(\kappa=0\), \(\epsilon=0\), or \(T_1=T_2\), consistent with the main-text discussion.

\section{Derivation of the heat current}
\label{app:heat_current_derivation}
\vspace{5mm}
This appendix derives the heat currents used in Sec.~\ref{sec:heat_entropy}. The heat current \(\dot Q_i\) is defined as the rate at which reservoir \(i\) injects energy into the configurational degrees of freedom,
\begin{equation}
	\dot Q_i
	=
	\int d\mathbf R\,
	\mathbf J_{i,\mathrm{ss}}(\mathbf R)
	\cdot
	\nabla_iU(\mathbf r_1,\mathbf r_2),
	\qquad
	i=1,2 .
	\label{eq:app_heat_definition}
\end{equation}
Here, \(\mathbf J_{i,\mathrm{ss}}\) denotes the two components of the full steady-state probability current associated with particle \(i\). With this convention, \(\dot Q_i>0\) means that reservoir \(i\) injects heat into the system.

The particle current is
\begin{equation}
	\mathbf J_{i,\mathrm{ss}}
	=
	-
	\boldsymbol{\mu}
	\left[
	P_{\mathrm{ss}}\nabla_iU
	+
	T_i\nabla_iP_{\mathrm{ss}}
	\right].
	\label{eq:app_heat_particle_current}
\end{equation}
Substituting Eq.~\eqref{eq:app_heat_particle_current} into Eq.~\eqref{eq:app_heat_definition} gives
\begin{equation}
	\dot Q_i
	=
	-
	\int d\mathbf R\,
	P_{\mathrm{ss}}
	\left(\nabla_iU\right)^{\mathrm T}
	\boldsymbol{\mu}^{\mathrm T}
	\nabla_iU
	-
	T_i
	\int d\mathbf R\,
	\left(\nabla_iP_{\mathrm{ss}}\right)^{\mathrm T}
	\boldsymbol{\mu}^{\mathrm T}
	\nabla_iU .
	\label{eq:app_heat_initial_expression}
\end{equation}
The first term is simplified by using \(\boldsymbol{\mu}=\mu_0(\mathbf I+\kappa\boldsymbol{\varepsilon})\). Since \(\boldsymbol{\varepsilon}^{\mathrm T}=-\boldsymbol{\varepsilon}\), the antisymmetric part does not contribute to the quadratic form,
\begin{equation}
	\left(\nabla_iU\right)^{\mathrm T}
	\boldsymbol{\mu}^{\mathrm T}
	\nabla_iU
	=
	\mu_0
	\left|
	\nabla_iU
	\right|^2 .
	\label{eq:app_heat_odd_quadratic_cancel}
\end{equation}
For the second term, integration by parts gives
\begin{equation}
	-
	T_i
	\int d\mathbf R\,
	\left(\nabla_iP_{\mathrm{ss}}\right)^{\mathrm T}
	\boldsymbol{\mu}^{\mathrm T}
	\nabla_iU
	=
	T_i
	\int d\mathbf R\,
	P_{\mathrm{ss}}\,
	\nabla_i\cdot
	\left(
	\boldsymbol{\mu}^{\mathrm T}
	\nabla_iU
	\right).
	\label{eq:app_heat_integration_by_parts}
\end{equation}
Since \(\nabla_i\nabla_iU=(k+\epsilon)\mathbf I\), and the trace of the antisymmetric part of \(\boldsymbol{\mu}\) vanishes, one obtains
\begin{equation}
	\nabla_i\cdot
	\left(
	\boldsymbol{\mu}^{\mathrm T}
	\nabla_iU
	\right)
	=
	2\mu_0(k+\epsilon).
	\label{eq:app_heat_divergence_term}
\end{equation}
Therefore
\begin{equation}
	\dot Q_i
	=
	-
	\mu_0
	\left\langle
	\left|
	\nabla_iU
	\right|^2
	\right\rangle_{\mathrm{ss}}
	+
	2\mu_0T_i(k+\epsilon).
	\label{eq:app_heat_average_formula}
\end{equation}
This form shows that the odd part of \(\boldsymbol{\mu}\) does not enter directly through the local quadratic dissipation. It enters indirectly through the steady-state covariance matrix.

The force gradients are
\begin{equation}
	\nabla_1U
	=
	(k+\epsilon)\mathbf r_1
	-
	\epsilon\mathbf r_2,
	\qquad
	\nabla_2U
	=
	(k+\epsilon)\mathbf r_2
	-
	\epsilon\mathbf r_1 .
	\label{eq:app_heat_force_gradients}
\end{equation}
Using the covariance entries from ~\ref{app:steady_state_derivation}, their steady-state variances are
\begin{equation}
	\left\langle
	\left|
	\nabla_1U
	\right|^2
	\right\rangle_{\mathrm{ss}}
	=
	2T_1(k+\epsilon)
	+
	\frac{
		\epsilon^2(k+\epsilon)(1+\kappa^2)(T_2-T_1)
	}
	{
		\Delta_\kappa
	},
	\label{eq:app_heat_force_variance_1}
\end{equation}
and
\begin{equation}
	\left\langle
	\left|
	\nabla_2U
	\right|^2
	\right\rangle_{\mathrm{ss}}
	=
	2T_2(k+\epsilon)
	-
	\frac{
		\epsilon^2(k+\epsilon)(1+\kappa^2)(T_2-T_1)
	}
	{
		\Delta_\kappa
	},
	\label{eq:app_heat_force_variance_2}
\end{equation}
where
\begin{equation}
	\Delta_\kappa
	=
	(k+\epsilon)^2+\kappa^2\epsilon^2 .
	\label{eq:app_heat_delta_kappa}
\end{equation}
Substituting Eqs.~\eqref{eq:app_heat_force_variance_1} and \eqref{eq:app_heat_force_variance_2} into Eq.~\eqref{eq:app_heat_average_formula} gives
\begin{equation}
	\dot Q_1
	=
	-
	\frac{
		\mu_0\epsilon^2(k+\epsilon)(1+\kappa^2)
	}
	{
		\Delta_\kappa
	}
	(T_2-T_1),
	\label{eq:app_heat_Q1}
\end{equation}
and
\begin{equation}
	\dot Q_2
	=
	\frac{
		\mu_0\epsilon^2(k+\epsilon)(1+\kappa^2)
	}
	{
		\Delta_\kappa
	}
	(T_2-T_1).
	\label{eq:app_heat_Q2}
\end{equation}
Thus \(\dot Q_1+\dot Q_2=0\), as required at steady state. For \(T_2>T_1\), the positive heat-transfer rate from the hot reservoir to the cold reservoir is therefore
\begin{equation}
	\mathcal J_Q
	=
	\dot Q_2
	=
	-\dot Q_1
	=
	\frac{
		\mu_0\epsilon^2(k+\epsilon)(1+\kappa^2)
	}
	{
		(k+\epsilon)^2+\kappa^2\epsilon^2
	}
	(T_2-T_1).
	\label{eq:app_heat_JQ}
\end{equation}

The dimensionless form follows from \(\tau_0=(\mu_0k)^{-1}\), \(\tilde\epsilon=\epsilon/k\), and \(\Theta=T_2/T_1\). Dividing Eq.~\eqref{eq:app_heat_JQ} by \(T_1/\tau_0\) yields
\begin{equation}
	\widetilde{\mathcal J}_Q
	=
	\frac{\mathcal J_Q}{T_1/\tau_0}
	=
	(\Theta-1)
	\frac{
		\tilde\epsilon^2(1+\tilde\epsilon)(1+\kappa^2)
	}
	{
		(1+\tilde\epsilon)^2+\kappa^2\tilde\epsilon^2
	} .
	\label{eq:app_heat_JQ_dimensionless}
\end{equation}
The corresponding enhancement relative to the even-mobility dimer is
\begin{equation}
	\mathcal R_Q
	=
	\frac{\mathcal J_Q(\kappa)}
	{\mathcal J_Q(0)}
	=
	\frac{
		(1+\kappa^2)(1+\tilde\epsilon)^2
	}
	{
		(1+\tilde\epsilon)^2+\kappa^2\tilde\epsilon^2
	} .
	\label{eq:app_heat_ratio}
\end{equation}
Both \(\widetilde{\mathcal J}_Q\) and \(\mathcal R_Q\) are even functions of \(\kappa\). Thus reversing the oddness reverses the chirality of the probability currents but leaves the scalar heat-transfer rate unchanged.

Finally, the entropy production rate is obtained from the reservoir entropy changes. With the convention of Eq.~\eqref{eq:app_heat_definition},
\begin{equation}
	\dot S_{\mathrm{tot}}
	=
	-
	\sum_{i=1}^{2}
	\frac{\dot Q_i}{T_i}.
	\label{eq:app_entropy_definition}
\end{equation}
Using \(\dot Q_2=\mathcal J_Q\) and \(\dot Q_1=-\mathcal J_Q\), we find
\begin{equation}
	\dot S_{\mathrm{tot}}
	=
	\mathcal J_Q
	\left(
	\frac{1}{T_1}
	-
	\frac{1}{T_2}
	\right).
	\label{eq:app_entropy_heat_current}
\end{equation}
Substitution of Eq.~\eqref{eq:app_heat_JQ} gives
\begin{equation}
	\dot S_{\mathrm{tot}}
	=
	\frac{
		\mu_0\epsilon^2(k+\epsilon)(1+\kappa^2)
	}
	{
		(k+\epsilon)^2+\kappa^2\epsilon^2
	}
	\frac{
		(T_2-T_1)^2
	}
	{
		T_1T_2
	} .
	\label{eq:app_entropy_dimensional}
\end{equation}
Equivalently,
\begin{equation}
	\tau_0\dot S_{\mathrm{tot}}
	=
	\frac{
		\tilde\epsilon^2(1+\tilde\epsilon)(1+\kappa^2)
	}
	{
		(1+\tilde\epsilon)^2+\kappa^2\tilde\epsilon^2
	}
	\frac{
		(\Theta-1)^2
	}
	{
		\Theta
	} .
	\label{eq:app_entropy_dimensionless}
\end{equation}
This expression is non-negative, vanishes for \(T_1=T_2\) or \(\epsilon=0\), and is invariant under \(\kappa\to-\kappa\).

\providecommand{\newblock}{}

\end{document}